\documentclass{article}

\usepackage{hyperref}
\usepackage{titlesec}
\usepackage[titletoc,toc,title]{appendix}
\usepackage{url}
\usepackage{graphicx}
\usepackage{float}
\usepackage{placeins} 

\begin{document}

\title{Quantifying MEV on L2s: A Study of Polygon, Arbitrum, and Optimism}
\author{\href{https://www.arthur.bagourd.com/}{Arthur Bagourd\footnote{\href{https://www.arthur.bagourd.com/}{https://www.arthur.bagourd.com/}}}  \ \& Luca Georges Francois\footnote{\href{https://github.com/0xpanoramix}{https://github.com/0xpanoramix}}}
\date{June 2023}
\maketitle

\textit{This work was financed by a research grant from Flashbots} \cite{Fb}.
\\

\textbf{Abstract}: This paper addresses the lack of research on quantifying Maximal Extractable Value (MEV) on Ethereum Layer 2 networks (L2s). Our findings reveal a substantial amount of MEV to be extracted on L2s, particularly on Polygon, with a lower bound of \$213 million surpassing previous estimates. We observe that the majority of detected MEV on L2s consists of arbitrage opportunities, as liquidations are rare. These results emphasize the need for continuous monitoring and analysis of MEV on L2s, promoting informed decision-making for network selection and highlighting the associated risks.

\section{Introduction}

\subsection{Background and motivation}

MEV is the profit extracted by searchers by impacting transaction ordering, inclusion or exclusion in a permissionless fashion. While most of the research on MEV has been focused on Ethereum Layer 1 (where most of the MEV occurs), we do not know how much MEV is actually happening on other chains. The latest research on the subject \cite{MEVonL2s} dates from a year ago and encountered several obstacles - as a matter of fact, quantifying MEV on L2s is no trivial task. Ethereum Layer 2 chains (L2s) are scaling solutions designed to improve the efficiency and performance of the Ethereum blockchain by processing transactions via sidechains while retaining the security and decentralization of the main Ethereum network. Quantifying the extracted and extractable MEV on L2s can give MEV searchers some insights on potential opportunities on these chains and give the public an idea of how much MEV actually happens on L2s and whether it needs to be addressed or not.

\subsection{Research goal and objectives}
The main goal of this research is to quantify the amount of MEV in Ethereum L2 systems and to understand its impact on the network. The specific objectives of this study are:

\begin{itemize}
    \item To port Flashbots' \texttt{mev-inspect-py} \cite{mip} software to L2s (Polygon, Optimism and Arbitrum)
    \item Quantifying historical MEV profit by block on these netweoks
    \item Classifying extracted MEV by domain (arbitrage vs liquidation)
    \item To compare the MEV on L2s to Ethereum
\end{itemize}

The focus here will be mainly on classic atomic arbitrages, rather than NFT sniping/arbitrage, token sniping at launch or other types of MEV.

\section{Literature Review}

\subsection{How does MEV work on L2s?}
Before understanding how MEV works on L2s, let us define what we mean by MEV. MEV "is a measure of the total value that can be extracted permissionlessly (i.e. without any special rights) from transaction ordering" \cite{Fbexplorer}. For a discussion on what can really be considered MEV or note see \cite{MEVDef}. In practice, it mostly consists of 2 types of transactions:
\begin{itemize}
    \item Arbitrages
    \item Liquidations
\end{itemize}
Arbitrage occurs when a searcher (a person/entity "searching" for MEV) spots an incoming transaction in the mempool that will change the reserves of a Decentralised Exchange (DEX), thus resulting in an arbitrage opportunity with another DEX. Arbitrages also contain sandwiches, which involve a front run leg before the target transaction, thus having it pushing the price even further, and closing after it with a backrun. 
\medskip

Liquidations extract MEV from selling off the collateral of a borrower that is used to secure a loan. This is generally because the oracle that determines the liquidation threshold will emit a new price that triggers it.
\medskip

On Ethereum, most of the extracted MEV goes through a block builder (be it Flashbots \cite{Fb} , builder0x69 \cite{builder0x69} ...). These are private RPCs that will receive the bundles from the searchers and build blocks by aggregating the bundles.

On other chains, there was no such system (Marlin \cite{Marlin} did not manage to become the Flashbots of Polygon) until recently (PFL \cite{PFL} recently received a lot of success on Polygon), the way MEV happens is by bribing the validators directly, either with off chain agreements or with on chain PGAs. Polygon is well known for its PGAs and transaction spamming. Looking at \href{https://polygonscan.com/txs?block=16675712}{this block} we can see that \textgreater 10\% of transactions are empty transactions from a single searcher. Indeed, the priority is given to gas ordering and then to time ordering, but because tx1 has been sent before tx2, assuming both have the same gas, this does not mean tx1 will be the first to be propagated through the network, hence the transaction spamming. On Polygon, when you see a target transaction in block $b$ and are sending a transaction to backrun it, it will only arrive in block $b+2$. This means that even if you are top of the block in $b+2$, if another searcher is sending a transaction in $b+1$ they can take the opportunity. Due to the gas being so cheap on Polygon, some searchers engage in "Optimistic" or "Probabilistic" arbitrage, by which they will send transactions that can exploit an arbitrage opportunity if there is one, and revert if there is none.

On Arbitrum and Optimism, there is only one sequencer and you cannot bribe it, transactions are ordered on a First Come First Served basis.
This means MEV is a latency game as to who is going to get the new state change and send arbitrage transactions the fastest. There is no transaction re-ordering being done by the searcher. Sandwiches and front-runs are impossible, only back-runs are possible. On Optimism, as gas is very cheap, searchers tend to run the whole logic on chain directly and run the WASM contract atomically in the same block.

\subsection{Previous research on MEV on L2s}

The main piece of research on the topic (\cite{MEVonL2s}) dates back to 2021. They were quite early in the field, so there was not much historical data to analyze (only 1 month of data for Optimism). Analyzing 1 month of data on Optimism they computed a total profit of \$34k. On Polygon, the total profit over was \$ 37M over 10 months. While they do not provide mean and average profit by day and block we can compute an average profit of \$0.12M/day and \$4.09 per transaction for Polygon and an average of \$1.13k/day and \$22.61 per transaction for Optimism. Now, what would be interesting is to see the median number, indeed as they noticed for Polygon, 16\% of the total profits were made on a single day. We can assume the distribution of MEV profits is very Pareto like in terms of days. They also noticed that a small number of tokens account for the majority of MEV profits. On Polygon: 63\% of the profits were done in MATIC, 20\% in ETH and 15\% in USDC. Significant events resulting in high volatility have a significant impact on the total arbitrage revenue.
\medskip

Marlin Protocol (\cite{Marlin}) has done extensive research on Polygon as well and they do publish numbers on their explorer. These numbers corroborate previous analysis and is more up to date with a \$45M total profit extracted to date. It does not show median or the distribution nor does it show the split by type of MEV.

\subsection{Obstacles encountered in previous research}
Let us highlight the challenges encountered in prior research related to the topic and assess the potential for their resolution.
\medskip

One of the obstacles faced in \cite{MEVonL2s} relates to Arbitrum and the fact that at the time it did not have a tracing API. Arbitrum now provides traces, which can be used to identify MEV transactions.
\medskip

Regarding Optimism, there was only 1 month of historical data available at the time, we now have 14 months of data available. 
\medskip

Due to Polygon's high frequency of blocks, it is difficult to conduct an analysis on historical data in a timely manner without any optimization, even when using logs (data that can be emitted during a smart contract's execution) and not traces (full records of the EVM execution), which is why the previous research relied on data provided by Marlin's Protocol.
\medskip

A final difficulty relates to the frequency of the token-to-USD prices, they only had access to hourly (or even worse daily) price data from Coingecko, which can lead to some transactions' profit being way under (and more rarely over) estimated due to the intra-day volatility of prices. Another issue is that using Coingecko's prices implies that one must have the token API id, which is a manual process, which resulted in them processing only the most common tokens.
We are using Uniswap Price Oracle \cite{uni} data applied to Uniswap on Arbitrum and Optimism as well as to QuickSwap on Polygon. This allows us to access block granular price data and gives the most precise profit estimation possible.

\section{Methodology}

\subsection{Overview of \texttt{mev-inspect-py} and its modifications for L2s}

\texttt{mev-inspect-py} \cite{mip} is a software developed by Flashbots \cite{Fb} to identify, classify and quantify MEV on Ethereum, by analyzing the transaction traces of each Ethereum block, classifying them as MEV transactions, and saving the results in a database. It works by fetching all the information from a block, classifying the traces for each transaction using the ABI (application binary interface) for all swap and liquidation-related functions, and then checking for different types of MEV such as arbitrages, liquidations, or cryptopunk snipes \cite{cryptopunk}.
\medskip

Marlin \cite{Marlin} developed a version that uses the logs instead of the traces, this means you look for the "swap" and "liquidation" events in the logs' topics instead of getting the whole trace and looking for the ABI. It runs faster and solves the issue of not having a tracing API available.
\medskip

We worked from Marlin's \cite{Marlin} version and adapted it for the different chains. We also added a module to analyze the profits of MEV transactions. To do so, we query the price of the token vs USDC on a Uniswap-V2 or Uniswap-V3 type DEX at the time of the transaction and get the equivalent USD profit. This allows to get a precision down to block granularity. Note that this gives the \textit{mark-to-market} PnL rather than the \textit{realized} PnL, as with the volatility of prices, a profit taken in a given token can quickly evaporate.
\medskip

Further analysis and aggregation is done by block/day/type of transaction, to get the average and median profits distributions, top tokens, etc.
\medskip

Note that we only analyze the most common MEV strategies: swaps and liquidations, we do not look at statistical arbitrages like strategies.

\section{Results}

\subsection{Extracted MEV on Polygon}

\subsubsection{Number of transactions}
First, let us look at the historical number of MEV transactions (by day):
\begin{figure}[H]
    \centering
    \includegraphics[width=1\textwidth]{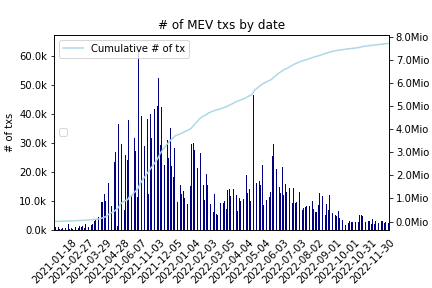}
    \caption{Number of transactions over time}
    \label{fig:poly-timeseries-of-tx-by-block}
\end{figure}

We have a total of 7.7 M transactions to date and we can also see a few days where it jumped massively.
\smallskip

Now, in terms of profit distribution by date:
\begin{figure}[H]
    \centering
    \includegraphics[width=1\textwidth]{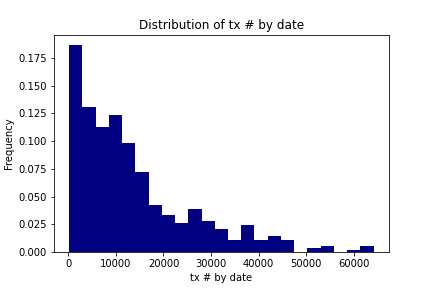}
    \caption{: Distribution of transactions by day}
    \label{fig:poly-distribution-of-tx-by-day}
\end{figure}

In terms of mean and median number of MEV transactions  for days and blocks when there is MEV:
\begin{center}
\begin{table}[htbp]
\centering
\begin{tabular}{|c|c|c|}
\hline
 & Median & Mean \\
\hline
Block & 1 & 1.93 \\
\hline
Date & 10071 & 13570.27 \\
\hline
\end{tabular}
   \caption{Median and mean number of transactions for days with MEV}
\end{table}
\end{center}

And looking back at all blocks and days from block 0:
\begin{center}
\begin{table}[htbp]
\centering
\begin{tabular}{|c|c|c|}
\hline
 & Median & Mean \\
\hline
Block & 0 & 0.26 \\
\hline
Date & 7439 & 11090.71 \\
\hline
\end{tabular}
   \caption{Median and mean number of transactions for all days}
\end{table}
\end{center}

\subsubsection{Extracted Profit}

Now onto the analysis of USD profits, let's look at the historical USD profit since inception:
\begin{figure}[H]
    \centering
    \includegraphics[width=1\textwidth]{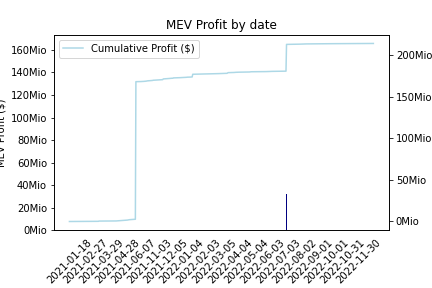}
    \caption{MEV profit over time}
    \label{fig:poly-distriubtion-of-profit-by-date}
\end{figure}
The total profit to date is \$213 M. We can see a massive \href{https://polygonscan.com/tx/0x653563c7c0977706406b8de0d946209ba5b09eca29d9cfd364004d658cf3f88d}{outlier MEV transaction}, let's remove it to get a better look:

\begin{figure}[H]
    \centering
    \includegraphics[width=1\textwidth]{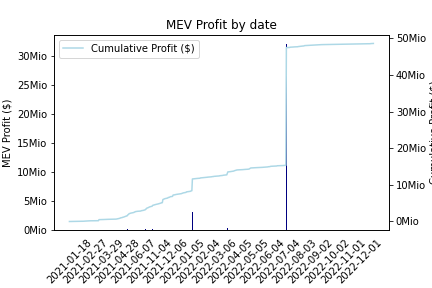}
    \caption{Distribution of MEV profit by date.}
    \label{fig:poly-distriubtion-of-profit-by-date}
\end{figure}

Let us look at the distribution of transactions with $\leq$ \$100 of profit, as this represents 99.99\% of all transactions:
\begin{figure}[H]
    \centering
    \includegraphics[width=1\textwidth]{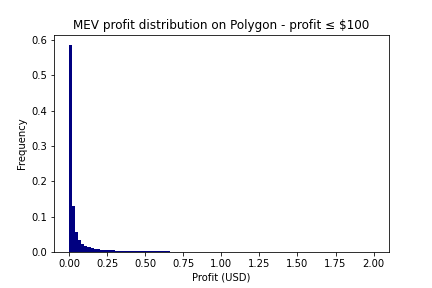}
    \caption{Distribution of MEV profit ($\leq$ \$100).}
    \label{fig:poly-distriubtion-of-profit-by-date}
\end{figure}
Though we can see the emergence of other buckets it looks like 90\% of the data still is in the bucket of transactions with $\leq$ \$1 of profit so let's look at that bucket:
\begin{figure}[H]
    \centering
    \includegraphics[width=1\textwidth]{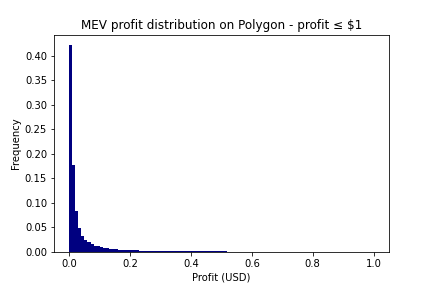}
    \caption{Distribution of MEV profit ($\leq$ \$1.0).}
    \label{fig:poly-distriubtion-of-profit-by-date}
\end{figure}

In terms of mean and median profit:
\begin{center}
\begin{table}[htbp]
\centering
\begin{tabular}{|l|c|c|}
\hline
\textit{Profit by} & Median & Mean \\
\hline
Block & 0 & 5.72 \\
\hline
Block with MEV & 0.02 & 53.70 \\
\hline
Date & 6,280.70 & 30,7644.98  \\
\hline
Date with MEV & 8,727.1 & 376,431.81 \\
\hline
MEV Tx & 0.011 & 27.74 \\
\hline
MEV Tx with strictly positive profit & 0.015 & 32.03 \\
\hline
\end{tabular}
    \caption{Median and mean profit (\$)}
\end{table}
\end{center}

\subsubsection{Top tokens}
What were the top tokens in which profits were taken and at what frequency?

\begin{center}
    \begin{table}[htbp]
    \centering
    \label{tab:token-profit}
    \begin{tabular}{l r  l r l}
        \hline
        Token & Count & Frequency (\%) \\
        \hline        
        matic-network & 3,853,613 & 52.32 \\
        usd-coin & 1,574,253 & 21.37 \\
        weth & 1,292,328 & 17.54 \\
        tether & 264,256 & 3.58 \\
        quick & 119,657 & 1.62 \\
        wrapped-bitcoin & 55,010 & 0.74 \\
        dai & 51,427 & 0.69 \\
        dfyn-network & 21,564 & 0.29 \\
        polydoge & 19,441 & 0.26 \\
        mmfinance & 16,417 & 0.22 \\
        \hline
    \end{tabular}
    \caption{Top 10 tokens profit was taken in}
\end{table}
\end{center}
\FloatBarrier

\subsubsection{Top transaction}
There is a one transaction that stands out:
\begin{itemize}
    \item hash: 0x653563c7c0977706406b8de0d946209ba5b09eca29d9cfd364004d658cf3f88d
    \item date: 17th of June 2021
    \item tokens involved: USDC and IRON (Titan)
    \item profit: \$165M (yes you read that correctly)
\end{itemize}
This transaction was missed by previous research, because those were using daily closing prices of Coingecko and here is the Close price from Coingecko on the 17th of June 2021, which is almost \$0.0:
\begin{figure}[H]
    \centering
    \includegraphics[width=1\textwidth]{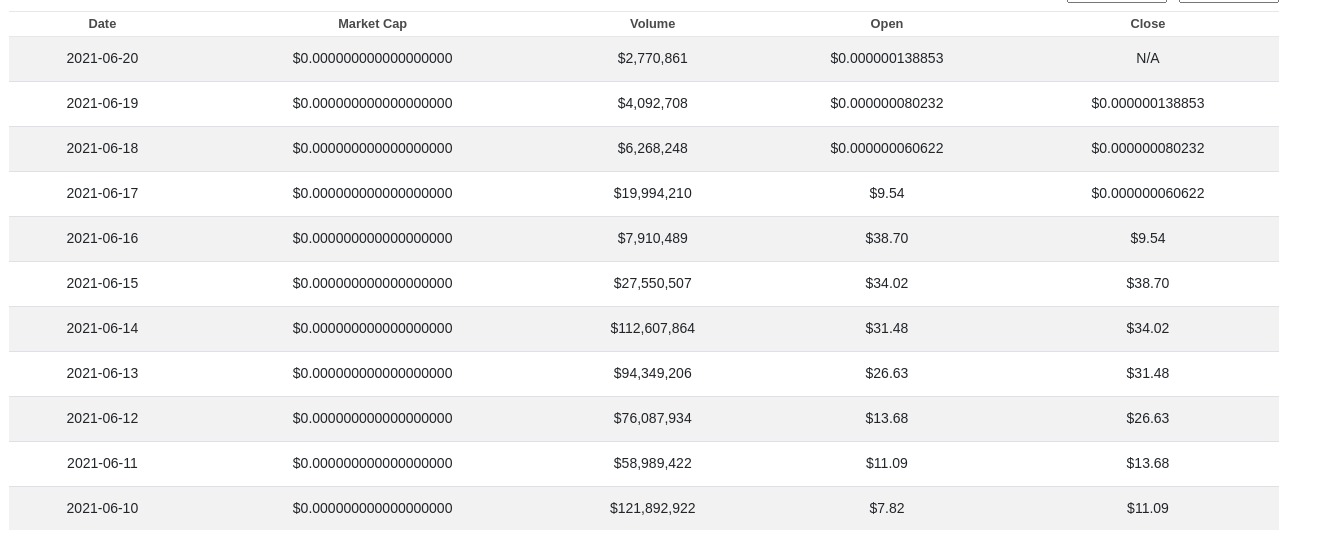}
    \caption{Coingecko prices for IRON (Titan) (see \href{https://www.coingecko.com/en/coins/iron-titanium-token/historical_data}{here})}
    \label{fig:poly-iron-price}
\end{figure}
It is really close to 0, while the trade happens at 01:41am when the price was closer to \$10.0, this really highlights the importance of using block granularity for prices, rather than daily prices, as well as the importance to get prices for all tokens rather than just the main ones (which is what one has to do using Coingecko as you need to fetch manually the Coingecko API id associated with the token).
\\

On the other hand, one should note that this is MtM PnL and not realised, so it is fully possible that by the time this position was exited the realised PnL was close to 0.

\newpage
\subsection{Extracted MEV on Arbitrum}

\subsubsection{Number of transactions}
First, let us look at the historical number of MEV transactions (by day):
\begin{figure}[H]
    \centering
    \includegraphics[width=1\textwidth]{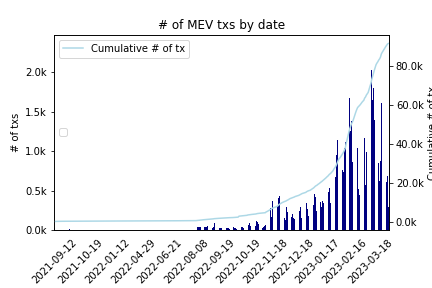}
    \caption{Number of transactions over time.}
    \label{fig:arbi-timeseries-of-tx-by-block}
\end{figure}

With a total of 87,706 MEV transactions. This seems very low and we do not have an explanation for that.
\medskip

Now, let's look at the distribution of the number of transactions per day:
\begin{figure}[H]
    \centering
    \includegraphics[width=1\textwidth]{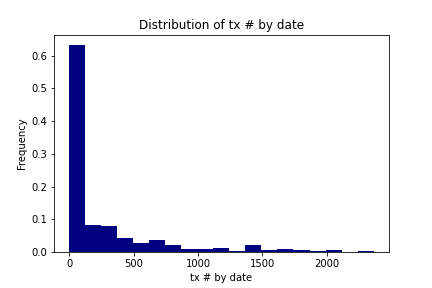}
    \caption{Distribution of transactions by day.}
    \label{fig:arbi-distribution-of-tx-by-day}
\end{figure}

In terms of mean and median number of MEV transactions for days and blocks when is MEV:
\begin{center}
\begin{table}[htbp]
\centering
\begin{tabular}{|c|c|c|}
\hline
 & Median & Mean \\
\hline
Block & 1 & 1.08 \\
\hline
Date & 30 & 233.88 \\
\hline
\end{tabular}
   \caption{Median and mean number of transactions for days with MEV}
\end{table}
\end{center}
And looking back at all blocks and days from block 0:
\begin{center}
\begin{table}[htbp]
\centering
\begin{tabular}{|c|c|c|}
\hline
 & Median & Mean \\
\hline
Block & 0 & 0.0011 \\
\hline
Date & 2 & 153.06 \\
\hline
\end{tabular}
   \caption{Median and mean number of transactions for all days}
\end{table}
\end{center}

\subsubsection{Extracted Profit}

Now onto the analysis of USD profits, let's look at the historical USD profit since inception:
\begin{figure}[H]
    \centering
    \includegraphics[width=1\textwidth]{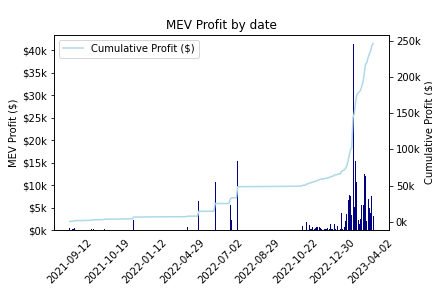}
    \caption{Extracted MEV profit by date.}
    \label{fig:arbi-distriubtion-of-profit-by-date}
\end{figure}
The total extracted MEV profit to date is of \$250k.
\\
Let's look at the distribution of transactions with $\leq$ \$100 of profit, as this represents 99.99\% of all transactions:
\begin{figure}[H]
    \centering
    \includegraphics[width=1\textwidth]{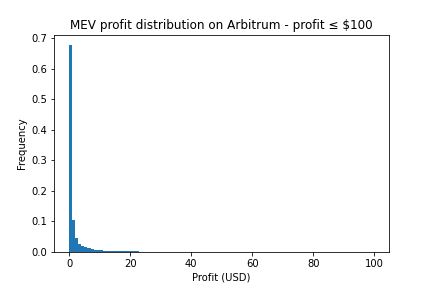}
    \caption{Distribution of MEV profit ($\leq$ \$100).}
    \label{fig:arbi-distriubtion-of-profit-by-tx-100}
\end{figure}
Though we can see the emergence of other buckets it looks like 90\% of the data still is in the bucket of transactions with $\leq$ \$10 of profit so let's look at that bucket:
\begin{figure}[H]
    \centering
    \includegraphics[width=1\textwidth]{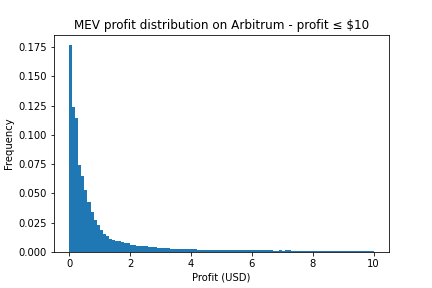}
    \caption{Distribution of MEV profit ($\leq$ \$10.0).}
    \label{fig:arbi-distriubtion-of-profit-by-tx-10}
\end{figure}
Let's look at the distribution of the profit for transactions earning less than \$1.0:
\begin{figure}[H]
    \centering
    \includegraphics[width=1\textwidth]{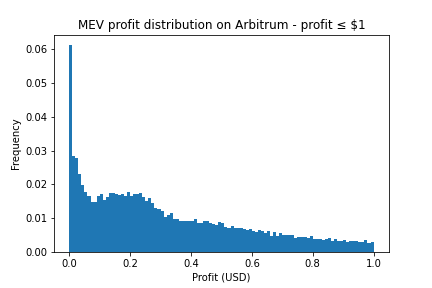}
    \caption{Distribution of MEV profit ($\leq$ \$1.0).}
    \label{fig:arbi-distriubtion-of-profit-by-tx-1}
\end{figure}
In terms of mean and median profit:
\begin{center}
\begin{table}[htbp]
    \centering
\begin{tabular}{|l|c|c|}
\hline
\textit{Profit by} & Median & Mean \\
\hline
Block & 0 & 0.007  \\
\hline
Block with MEV & 0.319 & 7.02  \\
\hline
Date & 6.06 & 987.05 \\
\hline
Date with MEV & 48.95 & 1,508.22 \\
\hline
MEV Tx  & 0.29 & 6.44 \\
\hline
MEV Tx with strictly positive profit & 0.48 & 8.02 \\
\hline

\end{tabular}
    \caption{Median and mean profit (\$)}
\end{table}
\end{center}

\subsubsection{Top tokens}
What were the top tokens in which profits were taken and at what frequency?
\begin{center}
    \begin{table}[htbp]
    \centering
    \label{tab:token-profit}
    \begin{tabular}{l r  l r l}
        \hline
        Token & Count & Frequency (\%) \\
        \hline        
        weth & 59,662 & 70.05 \\
        usd-coin & 14,581 & 17.12 \\
        radiant-capital & 2,062 & 2.42 \\
        jones-dao & 1,894 & 2.22 \\
        plutusdao & 1,334 & 1.57 \\
        magic & 1,137 & 1.33 \\
        tether & 1,021 & 1.20 \\
        sharky & 903 & 1.06 \\
        arbitrum & 872 & 1.02 \\
        swapfish & 519 & 0.61 \\
        \hline
    \end{tabular}
    \caption{Top 10 tokens profit was taken in}
\end{table}
\end{center}
\FloatBarrier

\newpage
\subsection{Extracted MEV on Optimism}

First, let us look at the historical number of MEV transactions (by day):
\begin{figure}[H]
    \centering
    \includegraphics[width=1\textwidth]{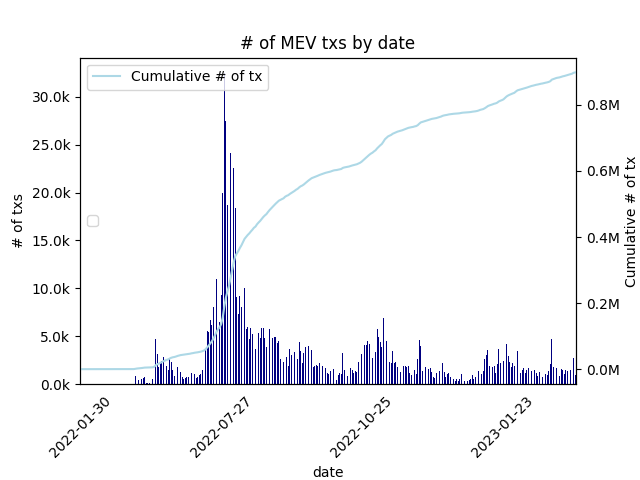}
    \caption{Number of transactions over time.}
    \label{fig:opti-timeseries-of-tx-by-block}
\end{figure}

There is a total of 900k MEV transactions with a peak at 30k in July 22.
\medskip

Now, let's look at the distribution of the number of transactions per day:
\begin{figure}[H]
    \centering
    \includegraphics[width=1\textwidth]{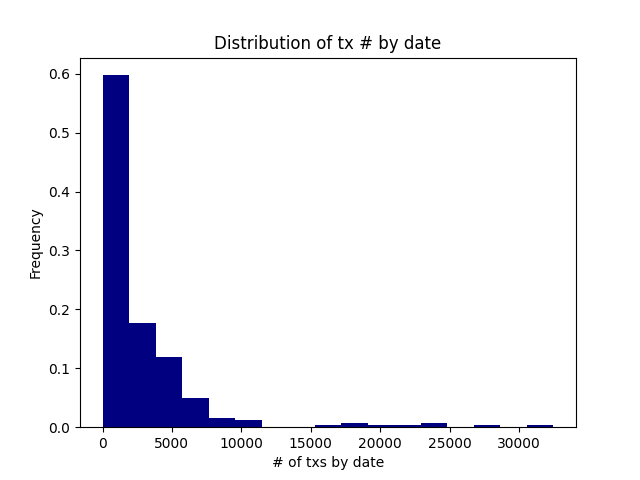}
    \caption{Distribution of transactions by day.}
    \label{fig:opti-distribution-of-tx-by-day}
\end{figure}
On 60\% of days there is less than 1.7k MEV tx. Note that this is without including all the days when only whitelisted accounts could deploy smart contracts on chain (up until December 21), so our results are not biased.

In terms of mean and median number of MEV transactions for days and blocks when is MEV:
\begin{center}
\begin{table}[htbp]
    \centering
\begin{tabular}{|c|c|c|}
\hline
 & Median & Mean \\
\hline
Block & 1 & 1.79 \\
\hline
Date & 1622 & 2824.75 \\
\hline
\end{tabular}
    \caption{Median and mean number of transactions for days with MEV}
\end{table}
\end{center}
And looking back at all blocks and days from block 0 (thus including all blocks before December 21):
\begin{center}
\begin{table}[htbp]
    \centering
\begin{tabular}{|c|c|c|}
\hline
 & Median & Mean \\
\hline
Block & 0 & 0.011 \\
\hline
Date & 1 & 1376.12 \\
\hline
\end{tabular}
    \caption{Median and mean number of transactions for all days}
\end{table}
\end{center}

\subsubsection{Extracted Profit}

Now onto the analysis of USD profits, let's look at the historical USD profit since inception:
\begin{figure}[H]
    \centering
    \includegraphics[width=1\textwidth]{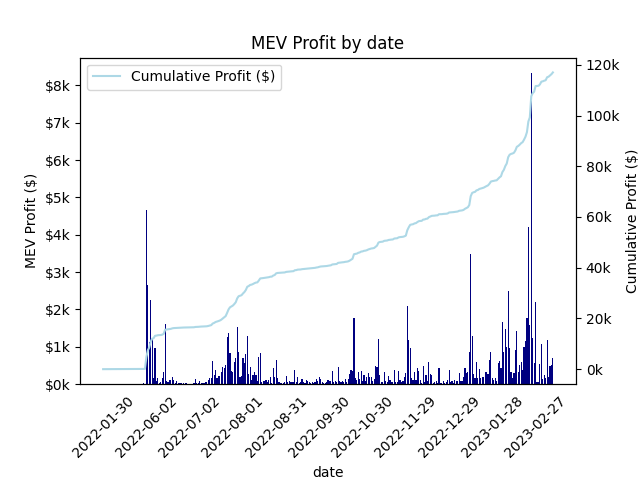}
    \caption{Extracted MEV profit by date.}
    \label{fig:opti-distriubtion-of-profit-by-date}
\end{figure}

We have a total MEV profit of \$ 120k to date.
\\

Let us look at the distribution of transactions with $\leq$ \$30 of profit:
\begin{figure}[H]
    \centering
    \includegraphics[width=1\textwidth]{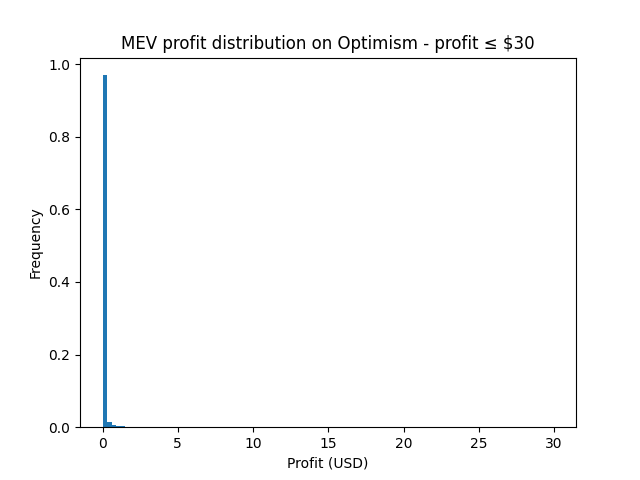}
    \caption{Distribution of MEV profit ($\leq$ \$100).}
    \label{fig:opti-distriubtion-of-profit-by-date}
\end{figure}
Though we can see the emergence of other buckets it looks like 90\% of the data still is in the bucket of transactions with $\leq$ \$1 of profit so let's look at that bucket:
\begin{figure}[H]
    \centering
    \includegraphics[width=1\textwidth]{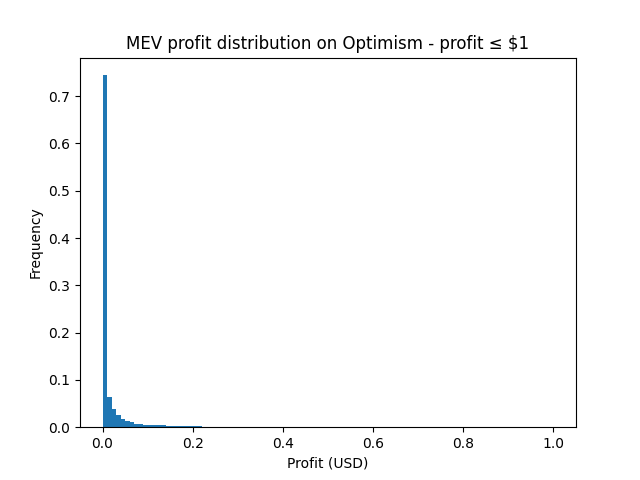}
    \caption{Distribution of MEV profit ($\leq$ \$1.0).}
    \label{fig:opti-distriubtion-of-profit-by-date}
\end{figure}

As for the distribution of MEV profit by day:
\begin{figure}[H]
    \centering
    \includegraphics[width=1\textwidth]{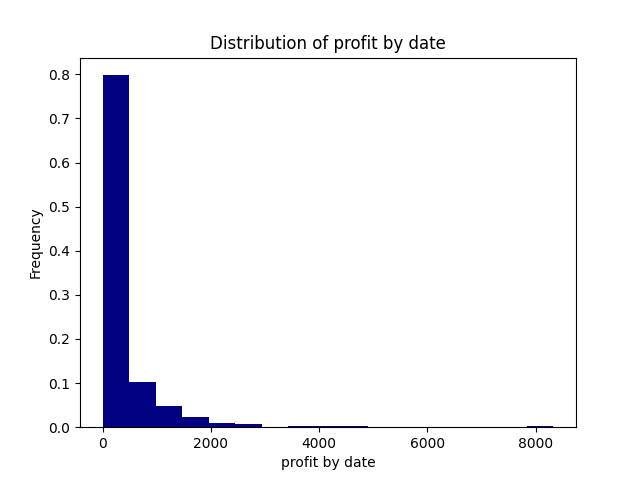}
    \caption{Distribution of MEV profit ($\leq$ \$1.0).}
    \label{fig:opti-distriubtion-of-profit-by-date}
\end{figure}
So on about 80\% of the days, the exctracted MEV profit is of less than \$500.

In terms of mean and median profit:
\begin{center}
\begin{table}[htbp]
    \centering
\begin{tabular}{|l|c|c|}
\hline
\textit{Profit by} & Median & Mean \\
\hline
Block & 0 & 0.0014  \\
\hline
Block with MEV & 0 & 0.24  \\
\hline
Date & 0 & 148.64 \\
\hline
Date with MEV & 126.36 & 373.74 \\
\hline
MEV Tx  & 0 & 0.14 \\
\hline
MEV Tx with strictly positive profit & 0.01 & 0.29 \\
\hline
\end{tabular}
    \caption{Median and mean profit (\$)}
\end{table}
\end{center}

\subsubsection{Top tokens}
What were the top tokens in which profits were taken and at what frequency?
\begin{center}
    \begin{table}[htbp]
    \centering
    \label{tab:token-profit}
    \begin{tabular}{l r  l r l}
        \hline
        Token & Count & Frequency (\%) \\
        \hline        
        usd-coin & 323,234 & 36.26 \\
        optimism & 206,601 & 23.17 \\
        weth & 145,701 & 16.34 \\
        velodrome-finance & 86,088 & 9.66 \\
        dai & 54,498 & 6.11 \\
        nusd & 26,426 & 2.96 \\
        mimatic & 19,497 & 2.19 \\
        havven & 12,889 & 1.45 \\
        frax & 6,186 & 0.69 \\
        liquity-usd & 5,399 & 0.61 \\
        \hline
    \end{tabular}
    \caption{Top 10 tokens profit was taken in}
\end{table}
\end{center}
\FloatBarrier

\newpage

\subsection{A comparison}

It is interesting to compare the total profit extracted on each chain. On Ethereum, we estimate it by adding the values provided by the Flashbots explorer pre merge (\$ 675) and post merge (\$74 M).
\begin{center}
\begin{table}[htbp]
    \centering
\begin{tabular}{|c|c|c|c|c|}
\hline
Polygon & Arbitrum & Optimism & Ethereum \\
\hline
213 & 0.25 & 0.12 & 749 \\
\hline
\end{tabular}
    \caption{Total profit (M \$) to date by chain}
\end{table}
\end{center}
Now, let us look at the mean and median profit aggregated at different granularities. It would be interesting to compare with Ethereum but these values are not provided by the Flashbots explorer.
\begin{center}
\begin{table}[htbp]
    \centering
\begin{tabular}{|l|c|c|c|}
\hline
\textit{Mean Profit by} & Polygon & Optimism &  Arbitrum \\
\hline
Block & 5.72 & 0.0014 & 0.007 \\
\hline
Block with MEV & 53.70 & 0.24 & 7.02 \\
\hline
Date & 30,7644.98 & 148.64 & 987.05 \\
\hline
Date with MEV & 376,431.81 & 373.74 & 1,508.22 \\
\hline
MEV Tx & 27.74 & 0.14 & 6.44 \\
\hline
MEV Tx with strictly positive profit & 32.03 & 0.29 & 8.02 \\
\hline
\end{tabular}
    \caption{Mean profit by chain}
\end{table}
\end{center}

\begin{center}
\begin{table}[htbp]
    \centering
    \begin{tabular}{|l|c|c|c|}
\hline
\textit{Median Profit by} & Polygon & Optimism & Arbitrum \\
\hline
Block & 0 & 0 & 0 \\
\hline
Block with MEV & 0.02 & 0 & 0.319 \\
\hline
Date & 6,280.70 & 0 & 6.06  \\
\hline
Date with MEV & 8,727.1 & 126.36 & 48.95 \\
\hline
MEV Tx & 0.011  & 0 & 0.29  \\
\hline
MEV Tx with strictly positive profit & 0.015 & 0.01 & 0.48  \\
\hline
\end{tabular}
    \caption{Median profit by chain}
\end{table}
\end{center}
\newpage

\subsection{Limitations}

A general limitation of \texttt{mev-inspect-py}\cite{mip} is that we are looking at blocks in batches so if an arbitrage spans different batches of blocks then it will not be detected.
\\

Also, when fetching prices we are looking at the price vs USDC, so if there is no pool of the token vs USDC, then we do not take into account that transaction. What should be done to improve this, is looking at the price of the token vs the chain's token (WETH on Ethereum, MATIC on Polygon etc), and then that token vs USDC.
\\
It is also important to note that we are looking at the market to market PnL, and not the realised one. We look at the price at a block granularity but a profit taken in a given token could quickly become 0 as the price of the said token goes to 0.
\\

For Polygon, it would also be great to find a way to differentiate the deterministic backrunning arbitrages from the opportunistic/probabilistic arbitrages.
\\

You may have noted the absence of classification of MEV by domain (in particular arbitrage vs liquidation), this is because liquidations are very scarce on these L2s, and 99.99\% of the detected MEV was arbitrages.

\section{Conclusion}

\subsection{Summary of the findings}
While MEV has been studied on Ethereum, there has not been a lot of studies on L2s. Our research finds that there is significant MEV extracted on other chains and in particular on Polygon. Previous estimates for Polygon (see Marlin \cite{Marlin} explorer) were at \$ 46 M, which is far from our finding of \$ 213 M, as they missed transactions. It is also important to note that our research is giving a lower bound of the actual extracted MEV, the methodology is not robust as seen in limitations and we are also missing the long tail MEV quantification (such as token sniping, cross-chain MEV, or other protocol exploits).

\subsection{Implications for MEV researchers and the public}
This shows that, despite the current market and the increasing competition, there still is still some MEV to be extracted on L2s. Of course this amount is way lower on Arbitrum and Optimism, but we do expect it to keep increasing as these 2 chains become more popular.

Moreover, our study highlights the importance of ongoing monitoring and analysis of MEV on L2s for the ecosystem. The findings of this research can increase awareness of the potential risks associated with MEV on L2s, and promote more informed decision-making in terms of selecting and using different blockchain networks as well as development of mechanisms to avoid transaction spamming (notably on Polygon). Additionally, quantifying MEV profit and transactions can also help provide a more accurate picture of the overall health and activity of different L2s, and help users and DApps in their decisions on chains selection.

\subsection{Recommendations for future research}
\subsubsection{Setup}
If you plan on running an empirical analysis you will make a lot of RPC queries, and these are the latency bottleneck. Before choosing an RPC provider, consider running a local node, this will speed up the RPC queries to less than 10ms each. If not, then keep in mind you will need to make a lot of requests and that RPC providers limit the rate of requests and the number of requests per day.

In terms of hardware setup, consider using an instance with 10 to 20 cores available.

\subsubsection{Potential Improvements}
 We are only getting prices vs USDC from Uniswap like pools, meaning that if a token does not have a pool with USDC on QuickSwap or Uniswap, we won't query its prices and we will skip it. The follow up improvements can be done:
 \begin{itemize}
     \item Querying prices vs the chain's token (MATIC for Polygon etc) and then the chain's token vs USDC
     \item if there is no price available, get the coingecko id of the token (can only be done manually) and query coingecko for that day and token, this will not be precise as you will get a daily price (00:00 UTC) but is probably still better than skipping the token
 \end{itemize}
\medskip

It would also be very interesting to see the evolution of MEV on Optimism with the recent Bedrock update \cite{bedrock}.

It is also worth noting that the \texttt{mev-inspect-py}\cite{mip} methodology is not the best, as it will miss some cases, one could improve the research by applying EigenPhi \cite{EigenPhi} methodology as it is more complete. We did not do it as we wanted to compare the total extracted MEV on each chain to the one extracted on Ethereum, so we had to keep the same methodology.

\newpage

\titleformat{\section}{\normalfont\Large\bfseries}{\appendixname~\thesection.}{1em}{}
\begin{appendices}

\section{Scaling}

\subsection{Description of the limitations of \texttt{mev-inspect-py}\cite{mip}  for long historical periods and our solutions}

For chains like Polygon, one of the arising issues is that the frequency of blocks is quite high (1 new block every 2s), which means that if you take more than that to analyze a block you will never catch up on real time. The same is true for Optimism and Arbitrum where transactions are added immediately, and then batched periodically to the L1.

We used logs instead of traces, as it speeds up the process.
\medskip

In order for us to avoid various limitations (rate-limit, request latency) due to the use of third party services to get the data we need, we opted for running our own archive nodes.
\medskip

\subsection{Computational environment of the analysis}

In the following subsections, we define the computational platform and the software environment used to perform the analysis.
We also present the methodology used during the analysis process itself, as we tweaked it to get the most out of the platform we had.

\subsubsection{Computational platform and Software Environment}

We ran both the archive nodes and the analysis process on the same machine, a Hetzner on-premise AX-101 server with a 16-cores AMD Ryzen 9 5950X CPU, 132 GB of RAM and approximately 6.9 TB of SSD storage.
The server is running Ubuntu 22.04.1 LTS (Linux Kernel version 5.15.0-72-generic).
In order to run \texttt{mev-inspect-py}\cite{mip}, we used minikube version v1.29.0.

\section{Running the analysis}
You can read the \href{https://github.com/IlluvatarEru/mev-inspect-py/blob/main/README.md}{README.md} on our repo to find instructions on how to run the analysis.

\end{appendices}

\end{document}